\documentstyle[preprint,prd,aps,epsf]{revtex}

\tightenlines

\begin{document}
\draft

\title{General Covariance in Quantum Gravity}

\author{Kirill~A.~Kazakov\thanks{Email address: $kirill@theor.phys.msu.su$}}

\address{Department of Theoretical Physics,
Physics Faculty,\\
Moscow State University,
$117234$, Moscow, Russian Federation}

\maketitle

\begin{abstract}
The question of general covariance in quantum gravity is considered in the
first post-Newtonian approximation. Transformation properties of observable
quantities under deformations of a reference frame, induced by variations of
the gauge conditions fixing general invariance, are determined. It is found
that the one-loop contributions violate the principle of general covariance,
in the sense that the quantities which are classically invariant under such
deformations take generally different values in different reference frames.
The relative value of this violation is of the order $1/N,$ where $N$ is the
number of particles in a gravitating body.
\end{abstract}
\pacs{04.60.D, 11.15.K, 11.10.L}

\section{Introduction}

Self-consistent quantization of gravitation is usually considered as a
high-energy problem. There is a number of models of quantum gravity none
of which has succeeded in reconciliation of renormalizability with basic
principles of the scattering theory, such as unitarity and causality.
It is important, however, that the low-energy properties of quantum gravity
are universal whatever the ultimate theory, in that they are determined
solely by the lowest order Einstein theory. Investigation of the low-energy
limit allows one to draw important conclusions about the synthesis of
quantum theory and gravitation.

One of the main principles underlying Einstein's general relativity
is the principle of general covariance. It states that arbitrary coordinate
transformations must leave the form of dynamical equations unchanged. Since
quantum theory deals with fields rather than coordinate transformations,
another formulation of this principle is appropriate for the purposes of
quantization: dynamical equations must be invariant under arbitrary
spacetime diffeomorphisms. This formulation reveals general relativity as a
gauge theory. From this point of view, any specific choice of coordinate
system, or more precisely, of reference frame, is equivalent to imposition
of an appropriate set of gauge conditions on the metric field, a change in
the gauge conditions being equivalent to a spacetime diffeomorphism.

In classical theory, there is no difference between the two points of view.
General covariance of the theory implies that it is diffeomorphism-invariant,
and vice versa. An important difference appears, however, in quantum theory.
While the notion of reference frame retains its essentially classical
content, components of the metric are promoted into operators, and so are
generators of the gauge transformations. Furthermore, the gauge conditions
become operator relations. Used in the classical theory as a means for
defining reference frames and their transformations, these notions thus
loose their direct interpretation in quantum domain. In this respect,
a natural question arises about relevant interpretation of the
above-mentioned operator relations, and their role in defining
reference frames in quantum gravity.

In connection with the above statement of the problem the following
circumstance should be emphasized. It is widely believed that the
characteristic length scale where quantum gravity effects come into
play is given by the Planck length
\begin{eqnarray}\label{lpscale}
l_{\rm P} = \sqrt{\frac{G\hbar}{c^3}}\ .
\end{eqnarray}
\noindent
The quantum gravitational corrections to the classical laws are thus
expected to be of the relative order $l^2_{\rm P}/l^2 = O(\hbar),$
where $l$ is the characteristic length of the problem under consideration.
Since these corrections reflect quantum properties of the spacetime itself,
one might doubt relevance of the notion of coordinate system in the
classical sense outlined above. One should note, however, that this
reasoning is based on the assumption that the Planck length is the
{\it only} scale of quantum gravitational effects. As far as pure gravity
is considered, this assumption is certainly true since $l^2_{\rm P}/\hbar$
is the only dimensional constant entering the Einstein action
\begin{eqnarray}&&\label{actionh}
S = - \frac{c^3}{16\pi G}{\displaystyle\int} d^4 x \sqrt{-g}R\,.
\end{eqnarray}
\noindent
In the presence of a matter field, however, another length scale appears
-- the gravitational radius
\begin{eqnarray}&&\label{rgscale}
r_g = \frac{2 G m}{c^2}\,,
\end{eqnarray}
\noindent
where $m$ is the mass of the field quanta. As a matter of fact,
along with terms proportional to $\hbar,$ the radiative gravitational
corrections also contain terms independent of the Planck constant
\cite{donoghue}\footnote{This property is specifically gravitational,
connected with the fact that the strength of this interaction is determined
by the masses of particles.}. Thus, the question whether the quantum gravity
effects appear already at the order $\hbar^0,$ or not, is the question of
{\it correspondence} between classical and quantum theories.

The problem of establishing the correct correspondence in quantum gravity
was considered in detail in Refs.~\cite{kazakov1,kazakov2}. It was shown,
in particular, that the approach using the S-matrix potential is inadequate
for this purpose, and the correct correspondence between classical and
quantum theories is to be established in terms of the effective (mean)
fields, rather than the S-matrix. This suggestion is
underlined by an observation that the $n$-loop radiative
contribution to the $n$th post-Newtonian correction to the mean
gravitational field of a body with mass $M,$ consisting of $N =
M/m$ elementary particles with mass $m,$ contains an extra factor
of $1/N^n$ in comparison with the corresponding tree contribution.
Thus, the effective gravitational field produced by the body turns
into the classical solution of the Einstein equations in the limit
$N \to \infty$ (and therefore, $M \to \infty$). It was also shown in
Ref.~\cite{kazakov1} that the S-matrix gravitational potential becomes
Newtonian in the same limit, hence, it fails to describe whatever
non-Newtonian interactions of {\it macroscopic} bodies.

An immediate consequence of this interpretation is that in the case of
{\it finite} $N,$ the loop corrections of the order $\hbar^0$ describe
deviations of the spacetime metric from classical solutions of the
Einstein equations, implying that (\ref{rgscale}) is the true scale of
quantum gravity effects as $\hbar\to 0.$

We can now reformulate the initial problem more precisely as follows.
On the one hand, the $\hbar^0$ radiative corrections are of the same
functional form as the post-Newtonian corrections predicted by classical
general relativity, and therefore must be treated on an equal footing
with the latter. On the other hand, their transformation under transitions
between different gauge conditions is expected to be more complicated.
The question is what the law of this transformation and its physical
interpretation are. This is actually the question of {\it general covariance
in quantum gravity.}\footnote{It is important that the restriction to zeroth
order in the Planck constant allows one to avoid, at least formally, the
difficult question about general physical interpretation of the quantum
corrections of higher orders.}

The aim of the present paper is to investigate this problem at the first
post-Newtonian approximation. The general approach used for this purpose
is outlined in Sec.~\ref{prelim}, where also the formulation of Einstein's
general covariance at the classical level is given in terms of quantum field
theory. The transformation law of the effective metric under variations of
the Feynman parameter (more generally, matrix of parameters) weighting the
gauge conditions in the action is established in Sec.~\ref{feynman}. It is
shown that these variations induce spacetime diffeomorphisms, and hence do
not change the values of observables. Comparison of this result with
classical theory is made. Transformation properties of the effective metric
under variations of the gauge conditions themselves are investigated in
Sec.~\ref{general}. This is done using the simplest model of the scalar
field minimally coupled to the gravitational field. The obtained results
are discussed in Sec.~\ref{conclude}.

Condensed notations of DeWitt \cite{dewitt1} are in force throughout
this paper. Also, right and left derivatives with respect to the
fields and the sources, respectively, are used. The dimensional
regularization of all divergent quantities is assumed.

\section{Preliminaries.}\label{prelim}

Before going into detailed discussion of the question of general covariance
in quantum gravity, let us describe the general setting we will be
working in.

\subsection{Frame of reference and interacting fields.}

First of all, we should set a frame of reference, {\it i.e.}, a system of
idealized reference bodies with respect to which the 4-position in spacetime
can be fixed. Let us assume, for definiteness, that the frame of reference
is realized by means of an appropriate distribution of electrically charged
matter. For simplicity, the energy-momentum of matter, as well as of the
electromagnetic field it produces, will be assumed sufficiently
small so as not to alter the gravitational field under consideration.
The 4-position in spacetime can be determined by exchanging
electromagnetic signals with a number of charged matter species.
The electric charge distributions $\sigma_a$ of the latter are
thus supposed to be in a one-to-one correspondence with the spacetime
coordinates $x_{\mu},$ $$\sigma_a\leftrightarrow x_{\mu}\,,$$
where index $a$ enumerates the species. The $\sigma_a(x)$ will be
assumed smooth scalar functions. Physical properties of the reference
frame are determined by the action $S_{\sigma}$ which specific form
is of no importance for us.

Next, let us consider a system of interacting gravitational and matter
fields. The latter are arbitrary species, bosons or fermions,
self-interacting or not, denoted collectively by $\phi_i,$ $i = 1,2,...,k.$
Dynamical variables of the gravitational field are $h_{\mu\nu} = g_{\mu\nu}
- \eta_{\mu\nu}.$ Dynamics of the system is described by the action
$S + S_{\phi},$ where $S_{\phi}$ is the matter action, and $S$ is
given\footnote{Our notation is $R_{\mu\nu} \equiv
R^{\alpha}_{~\mu\alpha\nu} =
\partial_{\alpha}\Gamma^{\alpha}_{\mu\nu} - \cdot\cdot\cdot,
~R \equiv R_{\mu\nu} g^{\mu\nu}, ~g\equiv \det g_{\mu\nu},
~g_{\mu\nu} = {\rm sgn}(+,-,-,-),$
$\eta_{\mu\nu} = {\rm diag}\{+1,-1,-1,-1\}.$
The Minkowski tensor $\eta$ is used to raise and lower tensor indices.
The units in which $\hbar = c = 16\pi G = 1$ are chosen in what follows.}
by Eq.~(\ref{actionh}).

The total action $S + S_{\phi} + S_{\sigma}$ is invariant under the gauge
transformations
\begin{eqnarray}\label{gaugesymm}
\delta h_{\mu\nu} &=&
\xi^{\alpha}\partial_{\alpha}h_{\mu\nu}
+ (\eta_{\mu\alpha} + h_{\mu\alpha})\partial_{\nu}\xi^{\alpha}
+ (\eta_{\nu\alpha} + h_{\nu\alpha})\partial_{\mu}\xi^{\alpha}
\equiv D_{\mu\nu}^{\alpha}\xi_{\alpha}\,,
\label{gaugesymmh}\\
\delta\phi_i &=&  D_i^{\alpha}\xi_{\alpha}\,,
\label{gaugesymmphi}\\
\delta\sigma_a &=& \sigma_{a,\alpha}\xi^{\alpha}
\end{eqnarray}
\noindent
The generators $D_{\mu\nu}, D_i$ span the closed algebra
\begin{eqnarray}\label{algebra}
D_{\mu\nu}^{\alpha,\sigma\lambda} D_{\sigma\lambda}^{\beta}
- D_{\mu\nu}^{\beta,\sigma\lambda} D_{\sigma\lambda}^{\alpha}
&=& f_{~~~\gamma}^{\alpha\beta} D_{\mu\nu}^{\gamma}\,,
\nonumber\\
 D^{\alpha,k}_i  D^{\beta}_k
-  D^{\beta,k}_i  D^{\alpha}_k
&=& f^{\alpha\beta}_{~~~\gamma}  D^{\gamma}_i\,,
\end{eqnarray}
where the "structure constants" $f^{\alpha\beta}_{~~~\gamma}$ are defined by
\begin{eqnarray}&&
f_{~~~\gamma}^{\alpha\beta}\xi_{\alpha}\eta_{\beta} =
\xi_{\alpha}\partial^{\alpha}\eta_{\gamma}
- \eta_{\alpha}\partial^{\alpha}\xi_{\gamma}\,.
\end{eqnarray}
\noindent
Let the gauge-fixing action be written in the form
\begin{eqnarray}\label{gaugefixpi}&&
S_{\rm gf} = \left(F_{\alpha}
- \frac{1}{2}\pi^{\beta}\zeta_{\beta\alpha}\right)\pi^{\alpha}\,,
\end{eqnarray}
\noindent
where $F_{\alpha}$ is a set of functions of the fields $h_{\mu\nu},$
fixing general invariance, $\pi^{\alpha}$ auxiliary fields introducing the
gauge, and $\zeta_{\alpha\beta}$ a non-degenerate matrix weighting
the functions $F_{\alpha};$ the particular choice
$\zeta_{\alpha\beta} = \xi\eta_{\alpha\beta}$ corresponds to the
well-known Feynman weighting of the gauge conditions.
Introducing the ghost fields $c_{\alpha},$ $\bar{c}^{\alpha},$
we write the Faddeev-Popov action \cite{faddeev}
\begin{eqnarray}\label{fp}
S_{FP} = S + S_{\phi} + S_{\sigma} + S_{\rm gf}
+ \bar{c}^{\beta}F_{\beta}^{,\mu\nu}D_{\mu\nu}^{\alpha}c_{\alpha}\,.
\end{eqnarray}
\noindent
$S_{FP}$ is invariant under the following Becchi-Rouet-Stora-Tyutin
(BRST) transformations \cite{brst}
\begin{eqnarray}\label{brst}
\delta h_{\mu\nu} &=& D_{\mu\nu}^{\alpha}c_{\alpha}\lambda\,,
\nonumber\\
\delta \phi_i &=&  D_i^{\alpha}c_{\alpha}\lambda\,,
\nonumber\\
\delta \sigma_a &=& \sigma_{a,\alpha}c^{\alpha}\lambda\,,
\nonumber\\
\delta c_{\gamma} &=& - \frac{1}{2}f^{\alpha\beta}_{~~~\gamma}
c_{\alpha}c_{\beta}\lambda\,,
\nonumber\\
\delta \bar{c}^{\alpha} &=& \pi^{\alpha}\lambda\,,
\nonumber\\
\delta\pi^{\alpha} &=& 0\,,
\nonumber
\end{eqnarray}
\noindent
where $\lambda$ is a constant anticommuting parameter.

\subsection{Generating functionals and Slavnov identities.}

The generating functional of Green functions has the form
\begin{eqnarray}\label{gener}&&
Z[J,K]
= {\displaystyle\int}d\Phi \exp\{i (\Sigma
+ \bar{\beta}^{\alpha}c_{\alpha} + \beta_{\alpha}\bar{c}^{\alpha}
+ t^{\mu\nu}h_{\mu\nu} + j^i\phi_i + s^a\sigma_a)\},
\end{eqnarray}
\noindent
where
\begin{eqnarray}&&
\Sigma = S_{FP}
+ k^{\mu\nu}D_{\mu\nu}^{\alpha}c_{\alpha}
+ q^i D_i^{\alpha}c_{\alpha}
+ r^a\sigma_{a,\alpha}c^{\alpha}
- \frac{l^{\gamma}}{2} f^{\alpha\beta}_{~~~\gamma}c_{\alpha}c_{\beta}
+ n_{\alpha}\pi^{\alpha}\,,
\nonumber
\end{eqnarray}
\{$t,$ $j,$ $s,$ $\bar{\beta},$ $\beta$\} $\equiv J$ ordinary
sources, and \{$k,$ $q,$ $r,$ $l,$ $n$\} $\equiv K$ the
BRST-transformation sources \cite{zinnjustin} for the fields
\{$h,$ $\phi,$ $\sigma,$ $c,$ $\bar{c}$\} $\equiv \Phi,$ respectively.

The functional $\Sigma$ can be written \cite{batvil}
$$\Sigma(\Phi,K)
= \Sigma^r\left(\Phi,\frac{\delta\Psi(\Phi,K)}{\delta\Phi}\right)\,, $$
where the reduced action
\begin{eqnarray}&&
\Sigma^r(\Phi,K) = S + S_{\phi} + S_{\sigma}
+ k^{\mu\nu}D_{\mu\nu}^{\alpha}c_{\alpha}
+ q^i D_i^{\alpha}c_{\alpha}
+ r^a\sigma_{a,\alpha}c^{\alpha}
- \frac{l^{\gamma}}{2} f^{\alpha\beta}_{~~~\gamma}c_{\alpha}c_{\beta}
+ n_{\alpha}\pi^{\alpha}\,,
\nonumber
\end{eqnarray}
\noindent
and the gauge fermion $$\Psi(\Phi,K) = K\Phi +
\bar{c}^{\alpha}\left(F_{\alpha}
- \frac{1}{2}\pi^{\beta}\zeta_{\beta\alpha}\right).$$
Hence, under infinitesimal variation of the gauge conditions, $\Sigma$
transforms as
\begin{eqnarray}&&
\delta\Sigma(\Phi,K) =
\frac{\delta\Delta\Psi(\Phi,K)}{\delta\Phi}
\frac{\delta\Sigma(\Phi,K)}{\delta K}\,.
\nonumber
\end{eqnarray}
\noindent
The corresponding variation of the generating functional
\begin{eqnarray}&&
\delta Z[J,K]
= i {\displaystyle\int}d\Phi \exp\{i (\Sigma + J\Phi)\}
\frac{\delta\Delta\Psi(\Phi,K)}{\delta\Phi}
\frac{\delta\Sigma(\Phi,K)}{\delta K}\,.
\nonumber
\end{eqnarray}
\noindent
Integrating by parts and omitting
$\delta^2 \Sigma/\delta K\delta\Phi\sim\delta(0)$ in the latter equation
gives
\begin{eqnarray}&&\label{zvar}
\delta Z[J,K]
= i J\frac{\delta}{\delta K}{\displaystyle\int}d\Phi
\Delta\Psi\exp\{i (\Sigma + J\Phi)\}\,.
\end{eqnarray}
\noindent
Since $\Sigma$ is invariant under the BRST transformation (\ref{brst}),
a BRST change of integration variables in Eq.~(\ref{gener}) gives the
Slavnov identity for the generating functional
\begin{eqnarray}\label{slav}
J\frac{\delta Z}{\delta K} = 0\,,
\end{eqnarray}
\noindent
which allows one to rewrite
Eq.~(\ref{zvar}) in terms of the generating functional of connected
Green functions, $W = - i\ln Z,$
\begin{eqnarray}&&\label{wvar}
\delta W[J,K]
=  J\frac{\delta}{\delta K}\langle \Delta\Psi\rangle \,,
\end{eqnarray}
\noindent
where $\langle X\rangle$ denotes the functional averaging
of $X$ \cite{tyutin}.

\subsection{General covariance at the tree level.}\label{covtree}

From the point of view of the general formalism outlined
in the preceding sections, the classical Einstein theory
corresponds to the tree approximation of the full quantum theory.
The tree contributions to the expectation values of field operators
coincide with the corresponding classical fields, and the effective
equations of motion
$$\left\langle \frac{\delta\Sigma}{\delta h_{\mu\nu}} \right\rangle
+ t^{\mu\nu} = 0\,,$$ expressing the translation invariance of the
functional integral measure, go over into the classical Einstein equations.
The results of the preceding section allow one, in particular, to
reestablish the general covariance of these equations.

As was mentioned in the Introduction, coordinate transformations are
replaced in the quantum theory by the field transformations. In particular,
transformations of the reference frame are represented by variations of
the fields $\sigma_i,$ induced by appropriate variations of the gauge
conditions. Namely, it follows from Eq.~(\ref{wvar}) at the
tree level that a gauge variation $\Delta F_{\alpha}$ induces the following
variations of the metric and reference fields
\begin{eqnarray}\label{vartreeg}
\delta g_{\mu\nu} &=& g_{\mu\nu,\alpha}\Xi^{\alpha} +
g_{\mu\alpha,\nu}\Xi^{\alpha}
+ g_{\nu\alpha,\mu}\Xi^{\alpha} \,, \\
\delta\sigma_a &=& \sigma_{a,\alpha}\Xi^{\alpha}\,,
\qquad \Xi_{\alpha} = \langle c_{\alpha}\Delta\Psi\rangle\,,
\label{vartreesigma}
\end{eqnarray}
\noindent
where $\Delta\Psi$ is the corresponding variation of the gauge fermion.

The functions $g_{\mu\nu},\sigma_a$ undergo the same variations
(\ref{vartreeg}), (\ref{vartreesigma}) under the spacetime diffeomorphism
$$x^{\mu} \to x^{\mu} + \delta x^{\mu},$$
with $\delta x^{\mu} = - \Xi^{\mu}\,.$
Let us consider any quantity entering the Einstein equations, for instance,
the scalar curvature $R.$ Under the above change of gauge conditions,
the tree value of $R$ measured at a point $\sigma^0$ of the reference
frame remains unchanged,
\begin{eqnarray}\label{example}
\delta R[g(x(\sigma^0))]
= \frac{\delta R}{\delta g_{\mu\nu}}\delta g_{\mu\nu}
+ \frac{\partial R}{\partial x^{\mu}}\delta x^{\mu}
= \frac{\partial R}{\partial x^{\alpha}}\Xi^{\alpha}
- \frac{\partial R}{\partial x^{\mu}}\Xi^{\mu} = 0\,.
\end{eqnarray}
\noindent
Analogously, the tree contribution to any tensor quantity
$O_{\mu\nu...}$ (or $O^{\mu\nu...}$), for instance, the
metric $g_{\mu\nu}$ itself, calculated at a fixed reference
point $\sigma^0,$ transforms covariantly (contravariantly),
as prescribed by the position of the tensor indices of the
corresponding operator. This is the manifestation of general covariance
of the classical Einstein theory in terms of quantum field theory.

\section{General covariance at the one-loop order.}

Let us now consider the transformation properties of the one-loop
contributions. As in Sec.~\ref{covtree},
we have to determine the effect of an arbitrary gauge variation
on the value of the effective metric, and also on the functions
$\sigma_a,$ {\it i.e.}, on the structure of the reference frame.
After that, the transformation law of observables, defined generally as
the diffeomorphism-invariant functions of the metric and reference fields,
can be determined in the way followed in Sec.~\ref{covtree}.
For definiteness, we will deal below with the
scalar curvature $R.$ Since we are interested in the one-loop contribution
to the first post-Newtonian correction, we can linearize $R$ in
$h_{\mu\nu}:$
\begin{eqnarray}\label{r}
R = \partial^{\mu}\partial^{\nu} h_{\mu\nu} - \Box h\,,
\qquad h\equiv \eta^{\mu\nu}h_{\mu\nu}\,.
\end{eqnarray}
\noindent

The transformation properties of $R$ under variations of
the weighting matrix $\zeta_{\alpha\beta}$ are considered in
Sec.~\ref{feynman}, and under variations of the gauge functions
$F_{\alpha}$ themselves in Sec.~\ref{general}.

\subsection{Dependence of observables on weighting parameters.}
\label{feynman}

Dependence of the effective fields on the weighting matrix
$\zeta_{\alpha\beta}$ can be determined in a quite general way
without specifying neither the gauge functions $F_{\alpha},$
nor the properties of gravitating matter fields. We begin with the
classical theory in Sec.~\ref{tree}, and then consider the one-loop
order in Sec.~\ref{oneloop}.

\subsubsection{The tree level.}\label{tree}

As we saw in Sec.~\ref{covtree}, arbitrary gauge variations
lead to the transformations of the classical fields, equivalent to the
spacetime diffeomorphisms, and thus do not affect the values of $R$.
In the particular case of variations of the weighting matrix, however,
not only $R,$ but also the effective fields themselves remain unchanged.
This means that the structure of reference frames in classical
theory is determined by the functions $F_{\alpha}$ only. As this
differs in the full quantum theory, a somewhat more detailed discussion
of this issue will be given in this Section.

To show the $\zeta_{\alpha\beta}$-independence of the classical metric,
let us first integrate the auxiliary fields $\pi^{\alpha}$ out of the
gauge-fixing action (\ref{gaugefixpi}),
\begin{eqnarray}\label{gaugefixp}&&
S_{\rm gf} \to
S^{\xi}_{\rm gf} = \frac{1}{2} F_{\alpha}\xi^{\alpha\beta} F_{\beta}\,,
\qquad \zeta_{\alpha\beta}\xi^{\beta\gamma} = \delta_{\alpha}^{\gamma}\,.
\end{eqnarray}
\noindent
The classical equations of motion thus become
\begin{eqnarray}\label{classeq}&&
\frac{\delta(S + S^{\xi}_{\rm gf})}{\delta g_{\mu\nu}} = - T^{\mu\nu},
\end{eqnarray}
\noindent
where $T^{\mu\nu}$ is the energy-momentum tensor of matter. Using
invariance of the action $S$ under the gauge transformations
\begin{eqnarray}\label{gaugesym}
\delta g_{\mu\nu} = \xi^{\alpha}\partial_{\alpha}g_{\mu\nu} +
g_{\mu\alpha}\partial_{\nu}\xi^{\alpha}
+ g_{\nu\alpha}\partial_{\mu}\xi^{\alpha} \equiv \nabla_{\mu}\xi_{\nu}\,,
\end{eqnarray}
\noindent
and taking into account the ``conservation law''
$\nabla_{\mu}T^{\mu\nu} = 0,$ one has from Eq.~(\ref{classeq})
\begin{eqnarray}&&\label{cderiv}
F_{\alpha}\xi^{\alpha\beta}
\frac{\delta F_{\beta}}{\delta g_{\mu\nu}(x)}
\nabla^{x}_{\mu}\delta(x - y) = 0\,.
\end{eqnarray}
\noindent
The matrix $M_{\beta}^{\nu}(x,y) =
\delta F_{\beta}/\delta g_{\mu\nu}(x)\nabla^{x}_{\mu}\delta(x - y)$ is
non-degenerate; its determinant $\Delta \equiv \det M_{\beta}^{\nu}(x,y)$
is just the Faddeev-Popov determinant, and therefore $\Delta \ne 0.$
Hence, one has from Eq.~(\ref{cderiv}) $F_{\alpha}\xi^{\alpha\beta} = 0,$
and, in view of non-degeneracy of $\xi^{\alpha\beta},$ $F_{\alpha} = 0.$
The classical metric is thus independent of the choice of the matrix
$\xi^{\alpha\beta},$ and in particular, of the replacements
$F_{\alpha} \to A_{\alpha}^{\beta}F_{\beta}.$ One can put this in another
way by saying that the weighting matrix has no geometrical meaning in
classical theory.

This differs, however, in quantum domain. The classical equations
(\ref{classeq}) are replaced in quantum theory by the effective equations
$$\frac{\delta \Gamma}{\delta g^{\rm eff}_{\mu\nu}} =
- T_{\rm eff}^{\mu\nu}\,,$$ where $\Gamma,$ $g^{\rm eff}_{\mu\nu},$ and
$T_{\rm eff}^{\mu\nu}$ are the effective action, metric, and
energy-momentum tensor of matter, respectively. In general, the fields
$g^{\rm eff}_{\mu\nu}$ do not satisfy the gauge conditions
$F_{\alpha} = 0,$ and moreover, depend on the choice of the weighting
matrix $\xi^{\alpha\beta};$  $\xi^{\alpha\beta}$-independence is
inherited only by the tree contribution.

Dependence on the choice of the weighting matrix generally represents an
excess of the gauge arbitrariness over the arbitrariness in the choice
of reference frame; it is therefore a potential source of ambiguity
in the values of observables. This dependence causes no gauge ambiguity
of observables only if it reduces to the symmetry transformations. In
other words, under (infinitesimal) variations of the matrix
$\xi^{\alpha\beta},$ the fields $g^{\rm eff}_{\mu\nu}$ must transform as
in Eq.~(\ref{gaugesym})
\begin{eqnarray}\label{gaugesymeff}
\delta g^{\rm eff}_{\mu\nu}
= \Xi^{\alpha}\partial_{\alpha}g^{\rm eff}_{\mu\nu} +
g^{\rm eff}_{\mu\alpha}\partial_{\nu}\Xi^{\alpha}
+ g^{\rm eff}_{\nu\alpha}\partial_{\mu}\Xi^{\alpha}\,,
\end{eqnarray}
\noindent
with some functions $\Xi^{\alpha}.$
It will be shown in the following Section that this is the case indeed.

\subsubsection{The one-loop level.}\label{oneloop}

\begin{figure}
\hspace{3cm}
\epsfxsize=7cm\epsfbox{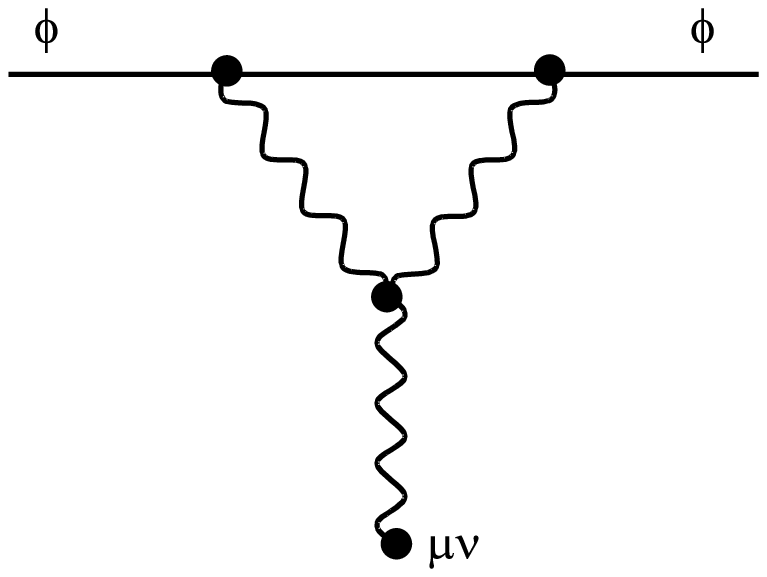}
\vspace{1cm}
\caption{The one-loop diagram
contributing to the first post-Newtonian correction. Wavy lines
represent gravitons, full lines massive particle.} \label{fig1}
\end{figure}

Let us now turn to examination of the gauge dependence of $\hbar^0$
loop contribution to the effective gravitational field.
This contribution comes from diagrams in which virtual propagation
of matter fields is near their mass shells, and is represented by terms
containing the root singularity with respect to the momentum transfer
between gravitational and matter fields. In the first
post-Newtonian approximation, the only diagram we need to consider is
the one-loop diagram pictured in Fig.~\ref{fig1}. As a simple
analysis shows, other one-loop diagrams do not contain
the root singularities, while the higher-loop diagrams are of
higher orders in the Newton constant.

\begin{figure}
\epsfbox[170 570 370 700]{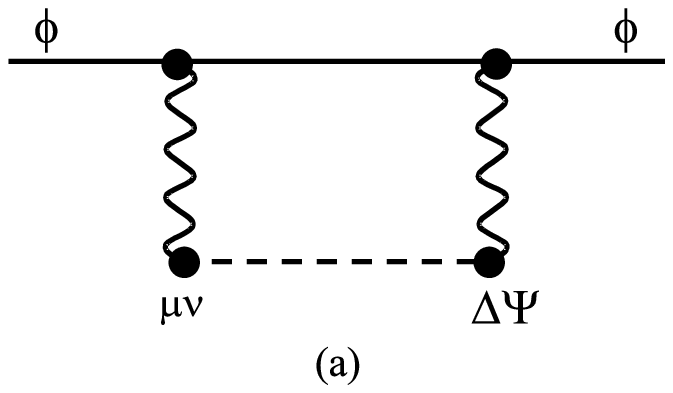}
\epsfbox[-80 460 140 580]{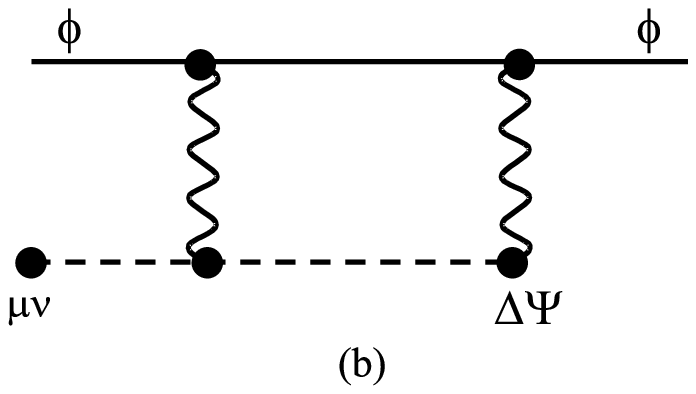}
\epsfbox[180 450 380 590]{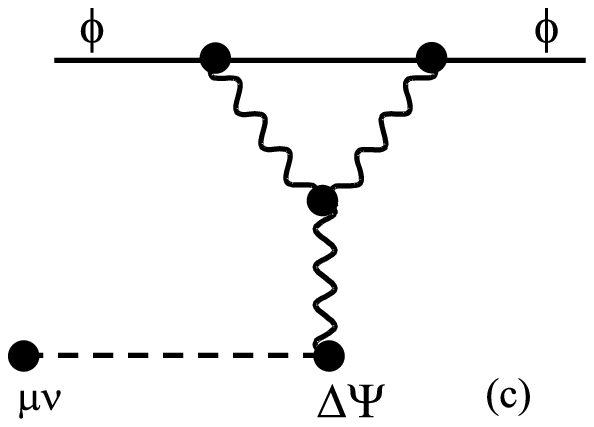}
\epsfxsize=8cm
\epsfbox[0 380 185 480]{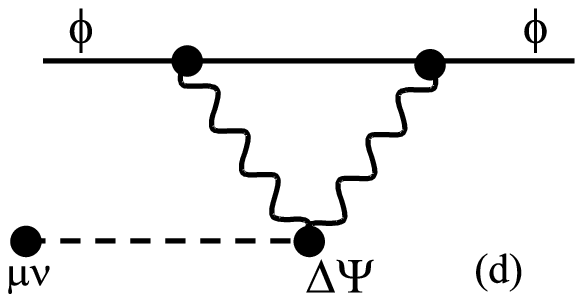}
\vspace{-5cm}
\caption{The one-loop diagrams
giving rise to the root singularity in the right hand side of
Eq.~(26). Dashed lines represent the Faddeev-Popov ghosts.}
\label{fig2}
\end{figure}

According to Eq.~(\ref{wvar}), under a variation $\Delta\Psi$ of the
gauge fermion, variation of the effective gravitational field
$$h^{\rm eff}_{\mu\nu} = \frac{\delta W}{\delta t^{\mu\nu}}$$ has the form
\begin{eqnarray}&&\label{hvar}
\delta h^{\rm eff}_{\mu\nu}
= \left( \frac{\delta}{\delta k^{\mu\nu}}\langle \Delta\Psi\rangle
+ j^i\frac{\delta^2}{\delta t^{\mu\nu}\delta q^i}
\langle \Delta\Psi\rangle \right)_{J\setminus j = 0 \atop K = 0}\,,
\end{eqnarray}
\noindent
where $J\setminus j$ means that the source $j$ is excluded from $J.$
We are interested presently in variations of the weighting matrix
$\xi^{\mu\nu},$ therefore,
$$\Delta\Psi(\Phi,K)
= - \frac{\bar{c}^{\alpha}}{2}\pi^{\beta}\Delta\zeta_{\beta\alpha}\,,$$
or, integrating $\pi^{\alpha}$ out,
\begin{eqnarray}&&\label{psivar}
\Delta\Psi(\Phi,K)
= \frac{\bar{c}^{\alpha}}{2}\zeta_{\alpha\beta}
\Delta\xi^{\beta\gamma}F_{\gamma}\,.
\end{eqnarray}
\noindent
According to general rules, in order to find the contribution of a
diagram with $n$ external $\phi$-lines, one has to take the $n$th
derivative of the right hand side of Eq.~(\ref{hvar}) with respect
to $j^i,$ multiply the result by the product of $n$ factors
$e_i(q^2 - m^2),$ where $e_i, q$ are the 4-momentum and polarization
of the external $\phi$-field quanta, and set $q^2 = m^2$ afterwards.
The second term on the right of Eq.~(\ref{hvar}) is proportional
to the source $j^i$ contracted with the vertex
$ D_i^{\alpha}c_{\alpha}.$ This term represents contribution
of the graviton propagators ending on the external matter lines.
Multiplied by $(q^2 - m^2),$ it gives rise to a non-zero value
as ${q^2\to m^2}$ only if the corresponding diagram is
one-particle-reducible with respect to the $\phi$-line, in which case
it describes the variation of $h^{\rm eff}_{\mu\nu}$ under the gauge
variation of external matter lines. It is well-known, however, that
$\phi$-operators must be renormalized\footnote{One might think that the
gauge dependence of the renormalization constants could spoil the
above derivation of Eq.~(\ref{wvar}). In fact, this equation holds
true for renormalized as well as unrenormalized quantities \cite{tyutin}.}
so as to cancel all the radiative corrections to the external
lines\footnote{The above discussion is nothing but the well-known reasoning
underlying the proof of gauge-independence of the $S$-matrix
\cite{tyutin2}.}. Therefore, this term can be omitted,
and Eq.~(\ref{hvar}) rewritten finally as
\begin{eqnarray}&&\label{hvar1}
\delta h^{\rm eff}_{\mu\nu}
= \frac{1}{2}\zeta_{\alpha\beta}
\Delta\xi^{\beta\gamma}
\left.\frac{\delta\langle\bar{c}^{\alpha} F_{\gamma}\rangle}
{\delta k^{\mu\nu}}\right|_{J\setminus j = 0 \atop K = 0}\,.
\end{eqnarray}
\noindent
The one-loop diagrams contributing to the right hand side of
Eq.~(\ref{hvar1}), giving rise to the root singularity, are pictured
in Fig.~\ref{fig2}. Let us consider the diagram of Fig.~\ref{fig2}(a) first.
It turns out that this diagram is actually free of the root singularity
despite the presence of the internal $\phi$-line. The rightmost vertex
in this diagram is generated by $\bar{c}^{\alpha}F^{(1)}_{\gamma},$ where
$F^{(1)}_{\gamma}$ denotes the linear part of $F_{\gamma}.$
The graviton propagator connecting this vertex to the $\phi$-line
can be expressed through the ghost propagator with the help of
the equation
\begin{eqnarray}\label{slavtree}
\xi^{\alpha\beta}F_{\beta}^{(1),\sigma\lambda} G_{\sigma\lambda\mu\nu} =
D^{(0)\beta}_{\mu\nu}\tilde{G}_{\beta}^{\alpha}\,,
\qquad D^{(0)\beta}_{\mu\nu}\equiv
\left.D^{\beta}_{\mu\nu}\right|_{h=0}\,,
\end{eqnarray}
\noindent
where $G_{\mu\nu\sigma\lambda},$ $ \tilde{G}_{\beta}^{\alpha}$ are the
graviton and ghost propagators defined by
\begin{eqnarray}\label{hpr}
\left.\frac{\delta^2 (S + S^{\xi}_{\rm gf})}
{\delta h_{\rho\tau}\delta h_{\mu\nu}}\right|_{h = 0}
G_{\mu\nu\sigma\lambda} = - \delta_{\sigma\lambda}^{\rho\tau}\,,
\qquad \delta_{\sigma\lambda}^{\rho\tau} = \frac{1}{2}
(\delta_{\sigma}^{\rho}\delta_{\lambda}^{\tau} +
\delta_{\sigma}^{\tau}\delta_{\lambda}^{\rho})\,, \nonumber
\end{eqnarray}
\noindent
and
\begin{eqnarray}&&
F_{\alpha}^{(1),\mu\nu}D^{(0)\beta}_{\mu\nu}\tilde{G}^{\gamma}_{\beta}
= - \delta_{\alpha}^{\gamma}\,, \nonumber
\end{eqnarray}
\noindent
respectively. Equation (\ref{slavtree}) is the Slavnov identity (\ref{slav})
at the tree level, differentiated twice with respect to
$t^{\mu\nu},$ $\beta_{\alpha}.$ Using this identity in the diagram
Fig.~\ref{fig2}(a) we see that the ghost propagator is attached to the
matter line through the generator $D^{(0)\alpha}_{\mu\nu}.$ On the
other hand, the action $S_{\phi}$ is invariant under the
gauge transformations (\ref{gaugesymm}),
\begin{eqnarray}&&
\frac{\delta S_{\phi}}{\delta\phi_i} D_i^{\alpha}
+ \frac{\delta S_{\phi}}{\delta h_{\mu\nu}}D^{\alpha}_{\mu\nu} = 0\,.
\end{eqnarray}
\noindent Differentiating this identity with respect to $\phi_k,$
setting $h_{\mu\nu} = 0,$ and taking into account that the
external $\phi$-lines are on the mass shell $$\left.\frac{\delta
S^{(2)}_{\phi}}{\delta\phi_i}\right|_{h=0} = 0\,,$$ where $S^{(2)}_{\phi}$
denotes the part of $S_{\phi}$ bilinear in $\phi,$ the $\phi^2 h$ vertex
can be rewritten
\begin{eqnarray}&&
\left.\frac{\delta^2 S^{(2)}_{\phi}}{\delta\phi_k\delta
h_{\mu\nu}}\right|_{h=0}D^{(0)\alpha}_{\mu\nu} = -
\left.\frac{\delta^2
S^{(2)}_{\phi}}{\delta\phi_i\delta\phi_k}\right|_{h=0} D_i^{\alpha}\,.
\end{eqnarray}
\noindent Thus, under contraction with the vertex factor, the
$\phi$-particle propagator, $G^{\phi}_{i k},$ satisfying
$$\left.\frac{\delta^2 S^{(2)}_{\phi}}{\delta\phi_i\delta\phi_k}\right|_{h=0}
G^{\phi}_{k l}
= - \delta_{l}^{i}\,,$$ cancels out
\begin{eqnarray}\label{cancel}
G^{\phi}_{k l}\left.\frac{\delta^2 S^{(2)}_{\phi}}{\delta\phi_k\delta
h_{\mu\nu}}\right|_{h=0}D^{(0)\alpha}_{\mu\nu} =  D_l^{\alpha}\,.
\end{eqnarray}
\noindent
We conclude that the $\hbar^0$ contribution of the diagram Fig.~\ref{fig2}(a)
is zero. As to the rest of diagrams, they are all proportional to the
generator $D^{(0)\alpha}_{\mu\nu}.$ Thus, the right hand side of
Eq.~(\ref{hvar1}) can be written
\begin{eqnarray}
\delta h^{\rm eff}_{\mu\nu} = D^{(0)\alpha}_{\mu\nu}\Xi_{\alpha}
+ O(\hbar)\,,
\qquad \Xi_{\alpha} = \frac{1}{2}\tilde{G}_{\alpha}^{\beta}
\zeta_{\beta\gamma}\Delta\xi^{\gamma\delta}\langle F_{\delta}\rangle \,.
\nonumber
\end{eqnarray}
\noindent
Since $\Xi_{\alpha}$ are of the order $G^2,$ one can also write, within
the accuracy of the first post-Newtonian approximation,
\begin{eqnarray}\label{mainaux}
\delta h^{\rm eff}_{\mu\nu} = D^{\alpha}_{\mu\nu}\Xi_{\alpha}\,,
\end{eqnarray}
\noindent
where $D^{\alpha}_{\mu\nu}$ are defined by Eq.~(\ref{gaugesymmh}) with
$h_{\mu\nu} \to h^{\rm eff}_{\mu\nu}\,.$

We thus see that under variations of the weighting matrix,
the effective metric does transform according to Eq.~(\ref{gaugesymeff}).
To determine the effect of these variations on the values of observables,
one has to find also the induced transformation of the reference frame,
{\it i.e.}, of the functions $\sigma_a.$ Obviously, the gauge variation of
$\sigma_a$'s is represented by the same set of diagrams pictured in
Fig.~\ref{fig2}(b,c,d),\footnote{The diagram of Fig.~\ref{fig2}(a)
does not contribute in this case.} with the only difference that the
leftmost vertex ($\mu\nu$) in these diagrams is now generated by
$\sigma_{a,\alpha}c^{\alpha}$ instead of $D^{\alpha}_{\mu\nu}c_{\alpha}.$
Thus, under variations of the weighting matrix, the functions $\sigma_a(x)$
transform according to
\begin{eqnarray}\label{mainaux1}
\delta \sigma_a = \sigma_{a,\alpha}\Xi^{\alpha}\,,
\end{eqnarray}
\noindent
where $\Xi^{\alpha}$'s are the same as in Eq.~(\ref{mainaux}).

Equations (\ref{mainaux}) and (\ref{mainaux1}) are of the same form
as Eqs.~(\ref{vartreeg}) and (\ref{vartreesigma}), respectively,
which implies that the value of any observable $O$ is invariant
under variations of the weighting matrix,
\begin{eqnarray}\label{example1}
\delta O[h^{\rm eff}(x(\sigma^0))]
= \frac{\delta O}{\delta h^{\rm eff}_{\mu\nu}}\delta h^{\rm eff}_{\mu\nu}
+ \frac{\partial O}{\partial x^{\mu}}\delta x^{\mu}
= \frac{\partial O}{\partial x^{\alpha}}\Xi^{\alpha}
- \frac{\partial O}{\partial x^{\mu}}\Xi^{\mu} = 0\,.
\end{eqnarray}
\noindent
In particular, $$\delta R[h^{\rm eff}(x(\sigma^0))] = 0.$$
Furthermore, any tensor quantity $O_{\alpha\beta...}$ (or $O^{\mu\nu...}$),
calculated at a fixed reference point $\sigma^0,$ transforms covariantly
(contravariantly), as prescribed by the position of the tensor indices of
the corresponding operator. This is in accord with the principle of general
covariance.

\subsection{Dependence of effective metric on the form of
$F_{\alpha}$'s.}\label{general}

Having established the general law of the effective metric transformation
under variations of the weighting matrix, let us turn to investigation of
the variations of the functions $F_{\alpha}$ themselves.

According to the general equation (\ref{hvar}), a variation
$\Delta F_{\alpha}$ induces the following variation in the
effective metric
\begin{eqnarray}&&\label{hvarf}
\delta h^{\rm eff}_{\mu\nu}
= \left.\frac{\delta\langle\bar{c}^{\alpha}\Delta F_{\alpha}\rangle}
{\delta k^{\mu\nu}}\right|_{J\setminus j = 0 \atop K = 0}\,.
\end{eqnarray}
\noindent
The general structure of diagrams representing the one-loop contribution to
the right hand side of this equation is the same as before and given by
Fig.~\ref{fig2}. Contribution of diagrams (b), (c), and (d) is again a
spacetime diffeomorphism. In the present case, however, diagram (a) gives
rise to a non-zero contribution already in the order $\hbar^0.$ Namely, it
is not difficult to show that the combination
$\Delta F^{(1),\mu\nu}_{\alpha} G_{\mu\nu\sigma\lambda}$ cannot be brought
to the form proportional to the generator $D^{(0)}.$ Note, first of all, that
the variation of $\Delta F^{(1),\mu\nu}_{\alpha} G_{\mu\nu\sigma\lambda}$
with respect to $\xi^{\alpha\beta}$ {\it is} proportional to $D^{(0)};$
in the highly condensed DeWitt's notation,
$$\delta (\Delta F^{(1)}_1 G) =
\Delta F^{(1)}_1 G (F^{(1)}_{1}\delta\xi F^{(1)}_1) G =
\Delta F^{(1)}_1 G F^{(1)}_1\delta\xi \zeta \tilde{G} D^{(0)},$$
where the Slavnov identity (\ref{slavtree}) has been used.
Hence, without changing the $\hbar^0$ part of diagram (a),
$\zeta_{\alpha\beta}$ can be set zero, in which case Eq.~(\ref{slavtree})
gives $F^{(1)}_1 G = 0.$ Suppose that
$\Delta F^{(1),\mu\nu}_{\alpha} G_{\mu\nu\sigma\lambda}
= X_{\alpha\beta}D^{(0)\beta}_{\sigma\lambda},$ or shorter,
$\Delta F^{(1)}_1 G = X D^{(0)},$ with some $X.$ Then one has
$0 = \Delta F^{(1)}_1 G F^{(1)}_1 = X F^{(1)}_1 D^{(0)}\equiv X M(h=0).$
Since the Faddeev-Popov determinant $\det M \ne 0,$ it follows that $X = 0.$
Thus, the argument used in the preceding section does not work, and
the question is whether contribution of the diagram (a) can be
actually represented in the form $D_{\mu\nu}^{\alpha}\Xi_{\alpha}.$

The answer to this question is negative, as an explicit calculation shows.
This will be demonstrated below for the simplest case of scalar matter
described by the action
\begin{eqnarray}&&\label{actionm}
S_{\phi} =  \frac{1}{2}{\displaystyle\int} d^4 x
\sqrt{-g}\left\{g^{\mu\nu}\partial_{\mu}\phi \partial_{\nu}\phi
- m^2\phi^2\right\},
\end{eqnarray}
\noindent
and linear gauge conditions
\begin{eqnarray}&&\label{gauge}
F_{\gamma} = \eta^{\mu\nu}\partial_{\mu}h_{\nu\gamma}
- \left(\frac{\varrho - 1}{\varrho - 2}\right)\partial_{\gamma}h\,,
\qquad h \equiv \eta^{\mu\nu}h_{\mu\nu}\,,
\qquad \zeta_{\alpha\beta} = 0\,,
\end{eqnarray}
\noindent
where $\varrho$ is an arbitrary parameter.
According to Eq.~(\ref{hvarf}), $\varrho$-dependence of the effective metric
is given by
\begin{eqnarray}&&\label{hvarfa}
\frac{\partial h^{\rm eff}_{\mu\nu}}{\partial\varrho}
= \left.\frac{\delta\langle\bar{c}^{\alpha}
\partial F_{\alpha}/\partial\varrho\rangle}
{\delta k^{\mu\nu}}\right|_{J\setminus j = 0 \atop K = 0}\,.
\end{eqnarray}
\noindent

\begin{figure}
\epsfxsize=17cm \epsfbox{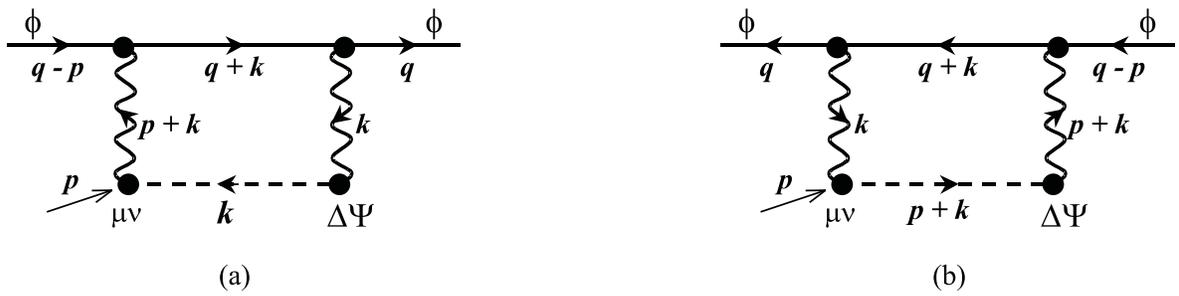}
\caption{Diagrams responsible for the nontrivial contribution
to the right hand side of Eq.~(34).} \label{fig3}
\end{figure}

There are two diagrams with the structure of Fig.~\ref{fig2}(a),
in which the scalar particle propagates in opposite directions.
They are represented in Fig.~\ref{fig3}. In fact, it is sufficient
to evaluate either of them. Indeed, these diagrams have the following
tensor structure $$a_1 q_{\mu}q_{\nu} + a_2 (p_{\mu}q_{\nu} + p_{\nu}q_{\mu})
+ a_3 p_{\mu}p_{\nu} + a_4\eta_{\mu\nu},$$ where $a_i, i = 1,...,4,$ are
some functions of $p^2.$ When transformed to the coordinate space,
the second and third terms become spacetime gradients, hence, they
can be written in the form $D_{\mu\nu}^{\alpha}\Xi_{\alpha}.$
As was discussed in the preceding sections, the terms of this type
respect general covariance, therefore, we can restrict ourselves to the
calculation of $a_1$ and $a_4$ only.
On the other hand, diagrams of Fig.~\ref{fig3} go over one into another
under the substitution $q \to p - q$ which leaves $a_1, a_4$ unchanged.

Calculation of the diagram \ref{fig3}(a) is somewhat easier. Its analytical
expression
\begin{eqnarray}&&\label{int}
I_{3(a)} =
- i\frac{\mu^{\epsilon}}{\sqrt{2\varepsilon_{\bf q} 2\varepsilon_{\bf q-p}}}
{\displaystyle\int}
\frac{d^{4-\epsilon} k}{(2\pi)^4}
\left\{
\frac{1}{2} W^{\alpha\beta\gamma\delta} (q_{\gamma} - p_{\gamma})
(k_{\delta} + q_{\delta})
- m^2 \frac{\eta^{\alpha\beta}}{2}
\right\}
\nonumber\\&&
\times G^{\phi}(q + k)\left\{
\frac{1}{2} W^{\rho\tau\sigma\lambda} q_{\sigma}
(k_{\lambda} + q_{\lambda}) - m^2 \frac{\eta^{\rho\tau}}{2}\right\}
G_{\rho\tau\pi\omega}(k)
\frac{\partial F^{\xi,\pi\omega}}{\partial\varrho}
\nonumber\\&&
\times\tilde{G}_{\xi}^{\zeta}(k)
\left\{- (k_{\zeta} + p_{\zeta}) \delta_{\mu\nu}^{\chi\theta}
+ \delta_{\zeta\mu}^{\chi\theta} k_{\nu}
+ \delta_{\zeta\nu}^{\chi\theta} k_{\mu} \right\}
G_{\chi\theta\alpha\beta}(k + p)\,,
\end{eqnarray}
\noindent
where the following notation is introduced:
\begin{eqnarray}
W^{\alpha\beta\gamma\delta} =
\eta^{\alpha\beta} \eta^{\gamma\delta}
- \eta^{\alpha\gamma} \eta^{\beta\delta}
- \eta^{\alpha\delta} \eta^{\beta\gamma}\,,
\nonumber
\end{eqnarray}
\noindent
$\mu$ -- arbitrary mass scale, $\varepsilon_{\bf q}
= \sqrt{{\bf q}^2 + m^2},$ and $\epsilon = 4 - d,$ $d$ being the
dimensionality of spacetime.
Explicit expressions for the propagators
\begin{eqnarray}
G_{\mu\nu\sigma\lambda} &=&
- \frac{W_{\mu\nu\sigma\lambda}}{\Box}
+ \varrho (\eta_{\mu\nu}\partial_{\sigma}\partial_{\lambda}
+ \eta_{\sigma\lambda}\partial_{\mu}\partial_{\nu})\frac{1}{\Box^2}
\nonumber\\
&-& (\eta_{\mu\sigma} \partial_{\nu} \partial_{\lambda}
+ \eta_{\mu\lambda} \partial_{\nu} \partial_{\sigma}
+ \eta_{\nu\sigma} \partial_{\mu} \partial_{\lambda}
+ \eta_{\nu\lambda} \partial_{\mu} \partial_{\sigma}) \frac{1}{\Box^2}
\nonumber\\
&-& (3\varrho^2 - 4\varrho)\partial_{\mu}\partial_{\nu}
\partial_{\sigma}\partial_{\lambda}\frac{1}{\Box^3}\,,
\label{hprop}\\
\tilde{G}^{\alpha}_{\beta} &=& - \frac{\delta^{\alpha}_{\beta}}{\Box}
+ \frac{\varrho}{2}\frac{\partial^{\alpha}\partial_{\beta}}{\Box^2}\,,
\nonumber\\
G^{\phi} &=&  \frac{1}{\Box + m^2}\,,
\nonumber
\end{eqnarray}
\noindent
Calculation of (\ref{int}) can be further simplified using the relation
$$\frac{\partial F_1}{\partial\varrho} G
= - F_1 \frac{\partial G}{\partial\varrho}\,,$$ which follows from
$F_1 G = 0,$ and noting that all gradient terms in the graviton
propagators, contracted with the $\phi^2 h$ vertices, can be omitted
(see Sec.~\ref{feynman}), i.e., only the first line in Eq.~(\ref{hprop})
actually contributes.

Let the equality of two functions up to a spacetime diffeomorphism
be denoted by ``$\sim$''. Then, performing tensor multiplications
in Eq.~(\ref{int}), and omitting terms proportional to $p_{\mu},$
one obtains
\begin{eqnarray}&&
I_{3(a)} \sim - i\frac{\mu^{\epsilon}}
{\sqrt{2\varepsilon_{\bf q} 2\varepsilon_{\bf q-p}}}
{\displaystyle\int}
\frac{d^{4-\epsilon} k}{(2\pi)^4}
\frac{1}{k^4}\frac{1}{(k + p)^2}\frac{1}{m^2 - (k + q)^2}
\nonumber\\&&
\left\{
\eta_{\mu\nu}\left[
\frac{\varrho }{2} (k^4 + 2 P Q) (P - Q) (Q - m^2)
\right.
\right.
\nonumber\\&&
\left.
\left.
+ \frac{\varrho }{2} k^2 P (P + Q) (Q - m^2))
- \varrho k^2 Q^2 (Q - m^2)
\right.
\right.
\nonumber\\&&
\left.
\left.
+ \left(\frac{\varrho}{4} p^2 (k^2 + 2 Q)
+ (k + p)^2 m^2 \right)(k^2 + P) (Q - m^2)
\right]
\right.
\nonumber\\&&
\left.
+ k_{\mu} k_{\nu} \left[
\varrho (P^2 + k^2 m^2) (Q - m^2)
+ 4 \varrho P Q m^2 - 3 \varrho P Q^2
\right.
\right.
\nonumber\\&&
\left.
\left.
+ \frac{\varrho}{2} p^2 (P - 2 Q) (Q - m^2)
+ 2 \varrho Q^2 (Q - m^2)
- \varrho P m^4
\right.
\right.
\nonumber\\&&
\left.
\left.
+ 2 (k + p)^2 (Q (Q - 2 m^2) - P (Q - m^2) + m^4)
\right]
\right.
\nonumber\\&&
\left.
+ 2 (k_{\mu} q_{\nu} + k_{\nu} q_{\mu})(k + p)^2
(Q - P) (Q - m^2)
\right.
\nonumber\\&&
\left.
- 2 q_{\mu} q_{\nu} (k + p)^2 (k^2 + P) (Q - m^2)
\right\}\,, \qquad Q\equiv (kq), \quad P\equiv (kp).
\end{eqnarray}
\noindent
Introducing the Schwinger parametrization of denominators
\begin{eqnarray}&&
\frac{1}{k^2} = - \int_{0}^{\infty} dy\exp\{y k^2\}\,,
\qquad\frac{1}{(k + p)^2} = - \int_{0}^{\infty} dx\exp\{x (k + p)^2\}\,,
\nonumber\\&&
\frac{1}{k^2 + 2 (kq)} = - \int_{0}^{\infty} dz \exp\{z [k^2 + 2 (kq)]\}\,,
\nonumber
\end{eqnarray}
one evaluates the loop integrals using
\begin{eqnarray}&&
\int~d^{d}k \exp\{ k^2 (x + y + z) + 2 k^{\mu} (x p_{\mu}
+ z q_{\mu}) + p^2 x \}
\nonumber\\&&
= i \left(\frac{\pi}{x + y + z}\right)^{d/2}
\exp\left\{\frac{p^2 x y - m^2 z^2}{x + y + z}\right\},
\nonumber
\end{eqnarray}
\begin{eqnarray}&&
\int~d^{d}k ~k_{\alpha}
\exp\{ k^2 (x + y + z) + 2 k^{\mu} (x p_{\mu}
+ z q_{\mu}) + p^2 x \} =
\nonumber\\&&
= i \left(\frac{\pi}{x + y + z}\right)^{d/2}
\exp\left\{\frac{p^2 x y - m^2 z^2}{x + y + z}\right\}
\left[- \frac{x p_{\alpha} + z q_{\alpha}}{x + y + z}\right],
\nonumber
\end{eqnarray}
\noindent
etc., up to six $k$-factors in the integrand.
This calculation can be automated to a considerable extent
with the help of the tensor package \cite{reduce} for the REDUCE system.
Changing the integration variables $(x,y,z)$ to $(t,u,v)$ via
$$x = \frac{t (1 + t + u) v}{m^2 (1 + \alpha t u)}\,,
\qquad y = \frac{u (1 + t + u) v}{m^2 (1 + \alpha t u)}\,,
\qquad z = \frac{ (1 + t + u) v}{m^2 (1 + \alpha t u)}\,,
\qquad \alpha \equiv - \frac{p^2}{m^2}\,,$$
integrating $v$ out, subtracting the ultraviolet divergence\footnote{
A technicality must be mentioned here. By itself, the diagram of
Fig.~\ref{fig1} is free of infrared divergences. As a result of the
BRST-operating with this diagram, however, some fictitious
infrared divergences are brought into individual diagrams representing
the right hand side of Eq.~(\ref{hvar}). This is because the vertex
$D^{\alpha}_{\mu\nu}C_{\alpha}$ contains the term
$C^{\alpha}\partial_{\alpha}h_{\mu\nu}$ in which the spacetime derivatives
act on the gravitational, rather than the ghost field. These divergences
occur as $u,t \to \infty.$ They are proportional to integer powers
of $p^2,$ and therefore do not interfere with the part containing the
root singularity. Since these divergences must eventually cancel in the
total sum in Eq.~(\ref{hvar}), they will be simply omitted in what follows.}
$$I_{3(a)}^{{\rm div}} = - \frac{1}{32\varepsilon_{\bf q}\pi^2 \epsilon}
\left(\frac{\mu}{m}\right)^{\epsilon}
\left[\frac{1}{3} q_{\mu} q_{\nu} + \eta_{\mu\nu}(p^2 - 2 m^2)
\frac{3\varrho - 2}{24} \right],$$
setting $\epsilon = 0$, omitting gradient terms, and retaining only
the $\hbar^0$-contribution, we obtain
\begin{eqnarray}\label{int2}
I^{\rm ren}_{3(a)}
&\equiv& (I_{3(a)} - I_{3(a)}^{{\rm div}})_{{\epsilon} \to 0}
\nonumber\\
&\sim&  \frac{m^2}{32\varepsilon_{\bf q}\pi^2}
\int_{0}^{\infty}\int_{0}^{\infty} du dt \left\{
\frac{\eta_{\mu\nu}\varrho}{H^3 \alpha}\left(
\frac{1}{D^2} - \frac{1}{2 D}\right)
\right.
\nonumber\\&+&
\left.
\frac{q_{\mu} q_{\nu}}{H^3 m^2}\left[
\frac{2 H^2}{D^2}\left(1 - \frac{1}{D}\right)
+ \frac{1}{\alpha}\left(\frac{4\varrho}{D^3} - \frac{11\varrho + 4}{D^2}
+ \frac{8\varrho + 4}{D}\right)
\right] \right\}\,,
\nonumber\\&&
D \equiv 1 + \alpha u t\,, \qquad H \equiv 1 + u + t\,.
\end{eqnarray}
\noindent
The root singularity in the right hand side of Eq.~(\ref{int2}) can be
extracted using the formulae derived in the appendix of Ref.~\cite{kazakov1}.
Denoting
$$\int_{0}^{\infty}\int_{0}^{\infty}
du dt~H^{- n} D^{- m} \equiv J_{nm} \,,$$
one has
\begin{eqnarray}
J^{{\rm root}}_{12} &=& \frac{\pi^2}{4\sqrt{\alpha}}\,,
~~J^{{\rm root}}_{13} = \frac{3\pi^2}{16\sqrt{\alpha}}\,,
\nonumber\\
J^{{\rm root}}_{31} &=& - \frac{\pi^2}{16}\sqrt{\alpha}\,,
~~J^{{\rm root}}_{32} = - \frac{3\pi^2}{32}\sqrt{\alpha}\,,
~~J^{{\rm root}}_{33} = - \frac{15\pi^2}{128}\sqrt{\alpha}\,.
\nonumber
\end{eqnarray}
\noindent
Substituting these into Eq.~(\ref{int2}) gives
\begin{eqnarray}&&\label{int3}
I^{\rm ren}_{3} = I^{\rm ren}_{3(a)} + I^{\rm ren}_{3(b)} \sim
2 I^{\rm ren}_{3(a)} \sim \frac{1}{256\varepsilon_{\bf q}\sqrt{\alpha}}
[q_{\mu} q_{\nu}(\varrho + 1) - \eta_{\mu\nu} m^2 \varrho].
\end{eqnarray}

Finally, restoring ordinary units, and going over to the coordinate
representation with the help of
\begin{eqnarray}
\int \frac{d^3{\bf p}}{(2\pi)^3}\frac{e^{i{\bf p x}}}{|{\bf p}|}
= \frac{1}{2\pi^2 r^2}\,,
\nonumber
\end{eqnarray}
\noindent
we obtain the following expression for the $\varrho$-derivative of
the $G^2$-order contribution to the effective metric
\begin{eqnarray}&&\label{main}
\frac{\partial h^{\rm eff}_{\mu\nu}}{\partial\varrho}
= \frac{\partial h^{\rm tree}_{\mu\nu}}{\partial\varrho}
+ \frac{\partial h^{\rm loop}_{\mu\nu}}{\partial\varrho}
\sim \frac{\partial h^{\rm loop}_{\mu\nu}}{\partial\varrho}
\sim \frac{G^2 m}{2 \varepsilon_{\bf q}c^2 r^2}
\left[\frac{q_{\mu} q_{\nu}}{c^2}(\varrho + 1)
- \eta_{\mu\nu} m^2 \varrho\right].
\end{eqnarray}
\noindent
The right hand side of this equation cannot be represented in the form
(\ref{gaugesymeff}). This result can be made more expressive by
calculating the $\varrho$-variation of the scalar curvature.
Setting ${\bf q} = 0$ for simplicity, we find that a variation
$\delta\varrho$ produces a non-zero variation of $R:$
\begin{eqnarray}
\delta R[h^{\rm eff}(x(\sigma^0))]
&=& \frac{\delta R}{\delta h^{\rm eff}_{\mu\nu}}
\left(\frac{\partial h^{\rm tree}_{\mu\nu}}{\partial\varrho}
+ \frac{\partial h^{\rm loop}_{\mu\nu}}{\partial\varrho}\right)\delta\varrho
+ \frac{\partial R}{\partial x^{\mu}}\delta x^{\mu}
\nonumber\\
&=& \frac{\partial R}{\partial x^{\alpha}}\Xi_{\rm tree}^{\alpha}
+ \left(\partial^{\mu}\partial^{\nu} - \eta^{\mu\nu}\Box\right)
\frac{h_{\mu\nu}^{\rm loop}}{\partial\varrho}\delta\varrho
- \frac{\partial R}{\partial x^{\mu}}\Xi_{\rm tree}^{\mu}
\nonumber\\
&=&
\partial_{i}\partial_{k}\frac{\partial h^{\rm loop}_{ik}}{\partial\varrho}
\delta\varrho
+ \Delta \frac{\partial h^{\rm loop}}{\partial\varrho}\delta\varrho
= \frac{G^2 m^2}{c^4r^4}(1 - 2\varrho)\delta\varrho\,,
\nonumber
\end{eqnarray}
\noindent
or\footnote{Another way to obtain this result is to introduce the sources
$t R$ and $k R_{,\alpha}c^{\alpha}$ for the scalar curvature and its
BRST-variation, respectively, into the generating functional (\ref{gener}),
instead of the corresponding sources for the metric. Then Eq.~(\ref{hvarfa})
is replaced by
\begin{eqnarray}\label{hvarfr}
\frac{\partial R^{\rm eff}}{\partial\varrho}
= \left.\frac{\delta\langle\bar{c}^{\alpha}
\partial F_{\alpha}/\partial\varrho\rangle}
{\delta k}\right|_{J\setminus j = 0 \atop K = 0}\,.
\end{eqnarray}
\noindent
At the second order in $G,$ the nontrivial contribution comes again from
the diagram of Fig.~\ref{fig2}(a) in which the lower left vertex is now
generated by $R_{,\alpha}c^{\alpha}.$ Thus, only the linear part of $R$
gives rise to a non-zero contribution to the right hand side of
Eq.~(\ref{hvarfr}). In other words,
$\delta R[h^{\rm eff}] = \delta R^{\rm eff}$, though generally
$R[h^{\rm eff}] \ne R^{\rm eff}.$}
\begin{eqnarray}\label{final}
\frac{\partial R[h^{\rm eff}(x(\sigma^0))]}{\partial\varrho}
= \frac{G^2 m^2}{c^4r^4}(1 - 2\varrho)\,.
\end{eqnarray}
\noindent
Equations (\ref{main}), (\ref{final}) express violation of
general covariance by the loop corrections.

\section{Discussion and conclusions}\label{conclude}

The results obtained in the preceding sections answer the general questions
stated in the Introduction. First of all, they establish general
transformation properties of observable quantities under
deformations of a reference frame, induced by variations of the gauge
conditions. Specifically, it was shown in Sec.~\ref{oneloop}
that although variations of the weighting matrix lead
to non-zero variations of the effective fields, the latter transform in such
a way that the observable quantities remain unchanged. Thus, the seemingly
wider freedom in the choice of gauge conditions at the quantum level
introduces no ambiguity into the values of the physical quantities.
In this sense, one can say that the physical properties of a given reference
frame are determined essentially by the form of the functions $F_{\alpha}$
only, just like in the classical theory.

Unlike the classical general relativity, however, observations of
a physical quantity in two reference frames defined by different
sets of functions $F_{\alpha}$ give generally different results.
For instance, spacetime curvature observed in a fixed reference
point $\sigma$ varies according to Eq.~(\ref{final}) under deformations
of the reference frame, induced by variations of the parameter $\varrho$
entering the gauge conditions (\ref{gauge}). The loop contributions thus
violate general covariance, depriving thereby the notion of spacetime
curvature of its absolute meaning, which is recovered only in the
macroscopic limit $N \to \infty,$ where $N$ is the number of elementary
particles producing the given gravitational field.

Thus, we arrive at the conclusion that {\it the principle of general
covariance is to be considered as approximate, valid only for the
description of macroscopic phenomena.}

Let us now discuss this issue from the practical point of view.
As was mentioned in the Introduction, from the point of view
of formal power expansion in $\hbar,$ the characteristic length scale
of quantum gravity effects is the same as in the classical Einstein theory.
Informally, the actual value of this scale depends on the physical
properties of a system under consideration. For fundamental elementary
particle such as the electron, $r_g$ is even smaller than the Planck length.
For the stars $r_g$ is measured by kilometers, but the quantum contribution
is highly suppressed in this case; in comparison with the classical (tree)
contribution, the loop contribution to the first post-Newtonian correction
to the field of a gravitating body contains the extra factor $1/N,$
where $N$ is the number of constituent particles. For the solar
gravitational field, for instance, this factor is of the order
$m_{proton}/M_{\odot} \approx 10^{-57}.$ However, this suppression is only
in force as long as interactions of the particles are relatively small.
This differs in a situation when the evolution of a system of
particles ends up with formation of the horizon. In this case,
interaction of particles in no way can be considered small.
From the point of view of an external observer, the number
$N$ is now irrelevant to the gravitational field of the collapsar
(this is a consequence of the ``no hair'' theorem). Made by the
infinite gravitational force indivisible, this object can be considered
as a ``particle'', {\it i.e.,} $N$ is to be set unity. As is well
known, black holes of certain types do behave like normal elementary
particles \cite{wilchek}. On the other hand, it should be emphasized
that from the point of view of the low-energy theory we work with,
the exact structure of the microscopic theory in which black hole is
embedded is of no importance. It is only important whether or not this
object can be described by a single quantum field. The loop contributions
to the gravitational field of black hole are thus of the same order of
magnitude as the ordinary post-Newtonian corrections predicted by classical
general relativity.

Calculation of the actual value of the one-loop contribution to the
effective gravitational field of black holes can be found in
Ref.~\cite{kazakov3}. Let us note in this connection that not only the
static gravitational field of black holes, but also emission of the
gravitational waves by the black hole binaries must be
affected by the quantum contributions. The LIGO and VIRGO \cite{ligo}
gravitational wave detectors, which are currently under
construction, will hopefully bring light into this issue.

{\bf Acknowledgments.}
I thank Drs. G.~A.~Sardanashvily and P.~I.~Pronin (Moscow State
University) for interesting discussions.

\pagebreak

\end{document}